\documentstyle[aps]{revtex}
\begin{document}
\title{Dynamic dislocation drag near to a point of martensitic transformation}
\author{V.N.Dumachev\thanks{%
E-mail: dumachev@edu.vrn.ru}}
\address{Voronezh Militia Institute, Russia}
\author{S.I.Moiseev and V.N.Nechaev}
\address{Voronezh State Technical University, Russia}
\maketitle

\begin{abstract}
The influence of coherent interface on dissipations of mechanical energy of
driven dislocations near to a point of martensite type phase transition is
considered. The expressions for dynamic braking of dislocations, owing to
losses of energy on excitation of deformations-type phase transition are
received.
\end{abstract}


Intensive experimental and theoretical study of dislocations dynamics [1-3]
has resulted in creation of correct classification of mechanisms of energy
dissipations of dislocation which drives in isotropic body [4-7]. In the
same time the questions connected to dynamic braking dislocations in a
crystal containing dot defects were considered [8,9]. It is known, that on
plastic and relaxations property of materials, essential influence renders
domain structure: its static and dynamic characteristics. Constructed in
works [10-11] the dynamic theory of coherent interphase boundaries allows at
a qualitatively new level to consider questions of interaction of elastic
fields created by dislocations and domain walls.

In present work the dynamic braking of dislocations in a crystal with
coherent interphase boundaries is investigated. As against earlier works on
this theme, self-consistent dynamic theory of interface here is used,
considering them as independent objects, which have of internal degrees of
freedom.

Let's proceed from the following expression for dissipations of energy D [12]

\begin{equation}
\overline{D}=2\pi \int \frac{d{\bf q}_{\Vert }}{\left( 2\pi \right) ^2}\int 
\frac{d\omega }{2\pi }\omega \left| f^{ext}({\bf q}_{\Vert }{\bf ,}\omega
)\right| ^2\text{Im }g({\bf q}_{\Vert }{\bf ,}\omega ),  \label{1}
\end{equation}

where

\begin{equation}
f^{ext}({\bf q}_{\Vert }{\bf ,}\omega )=-\left\{ \sigma _{ik}^{ext}\right\}
\left[ S_{ik}\right]  \label{2}
\end{equation}

- is configuration force acting on the part of moving dislocations in a
direction of boundary,

${\bf q}$ and $\omega $ wave vector and frequency of the elastic vibrations,

$q_{\Vert }^2=q_x^2+q_y^2$,

$\sigma _{ik}^{ext}$ - stress tensor created by an external source,

$\left[ S_{ik}\right] =1/2(m_iS_k+m_kS_i)$ plastic deformation tensor of
interface,

$m_i$ unit vector of the normal to the habit planes,

$S_k$ plastic displacement vector,

$\left\{ ...\right\} $ half-sum of the values of the braced quantity on both
sides of the boundary,

$\left[ ...\right] $ jump across the boundary.

Green function $g({\bf q}_{\Vert }{\bf ,}\omega )$ coherent interface we
receive by solving the dynamics equations of elastic theory together with
heat equation with the appropriate boundary conditions. (equality to zero of
configuration force (2) in each point of a border surface) [11] 
\begin{equation}
\rho \partial _{tt}^2u_i-\lambda _{iklm}\partial _k\left( \partial
_lu_m-s_{lm}\right) =0;  \label{3}
\end{equation}

\begin{equation}
\partial _tT-\chi \partial _{kk}^2T=%
%TCIMACRO{\dfrac{[H_{\circ }]}{C_v} }
%BeginExpansion
{\displaystyle {[H_{\circ }] \over C_v}}
%EndExpansion
\partial _t\zeta ({\bf r}_{\Vert },t)\delta (z);  \label{4}
\end{equation}

\begin{equation}
\lbrack S_{jk}]\{\sigma _{jk}\}+[H_{\circ }](T-T_{\circ })/T_{\circ }=0;
\label{5}
\end{equation}

where

$s_{ik}=\left[ S_{ik}\right] \delta (z)\zeta ({\bf r}_{\Vert },t)$ plastic
deformation tensor of body,

$\lambda _{iklm}=\lambda \delta _{ik}\delta _{lm}+\mu (\delta _{il}\delta
_{km}+\delta _{im}\delta _{kl})$ elastic constants tensor in isotropic
approximation,

$\lambda _{ik}^s=\lambda _{iklm}[S_{lm}]$,

$H_{\circ }$ enthalpy of the phase transformation,

$\zeta ({\bf r}_{\Vert },t)$ displacement of boundary along the normal
vector,

$C_v$ specific heat,

$\chi $ heat conductivity,

$T_{\circ }$ phase transitions temperature,

$\lambda ,\mu $ Lame's constants.

Equation (5) we shall write as

\[
\lbrack S_{ik}]\lambda _{iklm}\left( u_{lm}-\delta (z)\zeta ({\bf r}_{\Vert
},t)[S_{lm}]\right) _{z=0}+[H_{\circ }](T-T_{\circ })/T_{\circ }=0. 
\]

Solving the set of equations (3)-(5) in Fourier representation

\[
u_i\left( {\bf q},\omega \right) =-i\int 
%TCIMACRO{\dfrac{dq_z}{2\pi }}
%BeginExpansion
{\displaystyle {dq_z \over 2\pi }}
%EndExpansion
\left( G_{ik}\lambda _{kl}^sq_l\zeta ({\bf q}_{\Vert },\omega )\right) , 
\]

\[
T\left( {\bf q},\omega \right) =\frac{[H_{\circ }]}{C_v}\int 
%TCIMACRO{\dfrac{dq_z}{2\pi }}
%BeginExpansion
{\displaystyle {dq_z \over 2\pi }}
%EndExpansion
\frac{(-i\omega /\chi )}{q^2-i\omega /\chi }\zeta ({\bf q}_{\Vert },\omega
), 
\]

\[
0=iq_l\lambda _{lm}^su_m({\bf q},\omega )-\int 
%TCIMACRO{\dfrac{dq_z}{2\pi }}
%BeginExpansion
{\displaystyle {dq_z \over 2\pi }}
%EndExpansion
\zeta ({\bf q}_{\Vert },\omega )\lambda ^s+[H_{\circ }](T-T_{\circ
})/T_{\circ }, 
\]

for the Fourier-transform of function $\zeta ({\bf q}_{\Vert },\omega )$ we
find

\begin{equation}
\zeta ({\bf q}_{\Vert },\omega )\int \left( \lambda ^s-G\left( {\bf q}%
,\omega \right) +%
%TCIMACRO{\dfrac{[H_{\circ }]^2}{T_{\circ }C_v} }
%BeginExpansion
{\displaystyle {[H_{\circ }]^2 \over T_{\circ }C_v}}
%EndExpansion
%TCIMACRO{\dfrac{i\omega /\chi }{q^2-i\omega /\chi } }
%BeginExpansion
{\displaystyle {i\omega /\chi  \over q^2-i\omega /\chi }}
%EndExpansion
\right) 
%TCIMACRO{\dfrac{dq_z}{2\pi } }
%BeginExpansion
{\displaystyle {dq_z \over 2\pi }}
%EndExpansion
=0.  \label{6}
\end{equation}

Green function of equation (6) can be written as

\begin{equation}
g({\bf q}_{\Vert },\omega )=\frac 1{%
%TCIMACRO{\dint }
%BeginExpansion
\displaystyle \int 
%EndExpansion
\left( G(0,0,q_z,0)-G({\bf q},\omega )+%
%TCIMACRO{\dfrac{[H_{\circ }]^2}{T_{\circ }C_v} }
%BeginExpansion
{\displaystyle {[H_{\circ }]^2 \over T_{\circ }C_v}}
%EndExpansion
%TCIMACRO{\dfrac{i\omega /\chi }{q^2-i\omega /\chi } }
%BeginExpansion
{\displaystyle {i\omega /\chi  \over q^2-i\omega /\chi }}
%EndExpansion
\right) 
%TCIMACRO{\dfrac{dq_z}{2\pi } }
%BeginExpansion
{\displaystyle {dq_z \over 2\pi }}
%EndExpansion
}.  \label{7}
\end{equation}

Here

\[
\lambda ^s=[S_{ik}]\lambda _{iklm}[S_{lm}]=[S_{ik}]\lambda _{ik}^s, 
\]

\[
G=q_i\lambda _{ik}^sG_{kl}\lambda _{lm}^sq_m, 
\]

\begin{equation}
G_{ij}=%
%TCIMACRO{\dfrac 1{\mu (q^2-\omega ^2/c_t^2)} }
%BeginExpansion
{\displaystyle {1 \over \mu (q^2-\omega ^2/c_t^2)}}
%EndExpansion
\left( \delta _{ij}-%
%TCIMACRO{\dfrac{(1-\gamma ^2)q_iq_j}{q^2-\omega ^2/c_l^2} }
%BeginExpansion
{\displaystyle {(1-\gamma ^2)q_iq_j \over q^2-\omega ^2/c_l^2}}
%EndExpansion
\right)  \label{8}
\end{equation}

- Green function of dynamic equation of elastic theory,

$c_t,c_l$ sound velocity of transverse and longitudinal waves,

$\gamma =c_t/c_l$.

If boundary orientation is $S_i=\left( S_x,0,S_z\right) $, $n_i=(0,0,n_z)$
then

\begin{equation}
\lambda ^s=\mu (S_x^2+S_z^2/\gamma ^2),  \label{9}
\end{equation}

\begin{eqnarray}
G({\bf q},\omega ) &=&(\lambda q_j[S_{mm}]+2\mu q_i[S_{ij}])G_{jk}(\lambda
q_k[S_{mm}]+2\mu q_l[S_{lk}])  \label{10} \\
&=&\left( \lambda ^2q_jG_{jk}q_k+4\lambda \mu q_3G_{3k}q_k+4\mu
^2G_{33}q_3^2\right) S_3^2  \nonumber \\
&&+\left( \lambda \left( q_1G_{3k}q_k+q_3G_{1k}q_k\right) +2\mu \left[
G_{31}q_3^2+G_{33}q_1q_3\right] \right) 2\mu S_1S_3  \nonumber \\
&&+\left( 2G_{31}q_1q_3+G_{33}q_1^2+G_{11}q_3^2\right) \mu ^2S_1^2. 
\nonumber
\end{eqnarray}

Using (8) can be written (10) as

\begin{eqnarray}
G({\bf q},\omega ) &=&\left( \left[ \frac \lambda \mu +2%
%TCIMACRO{\dfrac{q_3^2}{q^2} }
%BeginExpansion
{\displaystyle {q_3^2 \over q^2}}
%EndExpansion
\right] ^2\frac{q^2\gamma ^2}{\kappa _l^2}+4%
%TCIMACRO{\dfrac{q_3^2}{q^2} }
%BeginExpansion
{\displaystyle {q_3^2 \over q^2}}
%EndExpansion
%TCIMACRO{\dfrac{q_{\Vert }^2}{\kappa _t^2} }
%BeginExpansion
{\displaystyle {q_{\Vert }^2 \over \kappa _t^2}}
%EndExpansion
\right) \mu S_3^2  \nonumber \\
&&+\left( \left[ 
%TCIMACRO{\dfrac \lambda \mu }
%BeginExpansion
{\displaystyle {\lambda \over \mu}}
%EndExpansion
+2%
%TCIMACRO{\dfrac{q_3^2}{q^2} }
%BeginExpansion
{\displaystyle {q_3^2 \over q^2}}
%EndExpansion
\right] \gamma ^2\frac{q_1q_3}{\kappa _l^2}+%
%TCIMACRO{\dfrac{q_3q_1}{q^2} }
%BeginExpansion
{\displaystyle {q_3q_1 \over q^2}}
%EndExpansion
%TCIMACRO{\dfrac{q_{\Vert }^2-q_3^2}{\kappa _t^2} }
%BeginExpansion
{\displaystyle {q_{\Vert }^2-q_3^2 \over \kappa _t^2}}
%EndExpansion
\right) 4\mu S_1S_3  \nonumber \\
&&+\left( \gamma ^2%
%TCIMACRO{\dfrac{4q_1^2q_3^2/q^2}{\kappa _l^2} }
%BeginExpansion
{\displaystyle {4q_1^2q_3^2/q^2 \over \kappa _l^2}}
%EndExpansion
+%
%TCIMACRO{\dfrac{q_1^2+q_3^2-4q_1^2q_3^2/q^2}{\kappa _t^2} }
%BeginExpansion
{\displaystyle {q_1^2+q_3^2-4q_1^2q_3^2/q^2 \over \kappa _t^2}}
%EndExpansion
\right) \mu S_1^2,  \nonumber
\end{eqnarray}

corresponding for the Green function (7) we readily get

\begin{equation}
g^{-1}({\bf q}_{\Vert },\omega )=\mu S^2%
%TCIMACRO{\dint }
%BeginExpansion
\displaystyle \int 
%EndExpansion
\left( 
\begin{array}{c}
\left( 4 
%TCIMACRO{
%\dfrac{q_{\Vert }^2\left( q_{\Vert }^2-\omega ^2/c_t^2\right) }{\left( q^2-\omega ^2/c_t^2\right) \left( q^2-\omega ^2/c_l^2\right) } }
%BeginExpansion
{\displaystyle {q_{\Vert }^2\left( q_{\Vert }^2-\omega ^2/c_t^2\right)  \over \left( q^2-\omega ^2/c_t^2\right) \left( q^2-\omega ^2/c_l^2\right) }}
%EndExpansion
\left( 1-\gamma ^2\right) -%
%TCIMACRO{\dfrac{\omega ^2/c_t^2}{\left( q^2-\omega ^2/c_l^2\right) } }
%BeginExpansion
{\displaystyle {\omega ^2/c_t^2 \over \left( q^2-\omega ^2/c_l^2\right) }}
%EndExpansion
\right) 
%TCIMACRO{\dfrac{S_3^2}{S^2} }
%BeginExpansion
{\displaystyle {S_3^2 \over S^2}}
%EndExpansion
\\ 
-\left( \left[ 
%TCIMACRO{\dfrac \lambda \mu }
%BeginExpansion
{\displaystyle {\lambda \over \mu}}
%EndExpansion
+2%
%TCIMACRO{\dfrac{q_3^2}{q^2} }
%BeginExpansion
{\displaystyle {q_3^2 \over q^2}}
%EndExpansion
\right] \gamma ^2%
%TCIMACRO{\dfrac{q_1q_3}{\left( q^2-\omega ^2/c_l^2\right) } }
%BeginExpansion
{\displaystyle {q_1q_3 \over \left( q^2-\omega ^2/c_l^2\right) }}
%EndExpansion
+%
%TCIMACRO{\dfrac{q_3q_1}{q^2} }
%BeginExpansion
{\displaystyle {q_3q_1 \over q^2}}
%EndExpansion
%TCIMACRO{\dfrac{q_{\Vert }^2-q_3^2}{\left( q^2-\omega ^2/c_t^2\right) } }
%BeginExpansion
{\displaystyle {q_{\Vert }^2-q_3^2 \over \left( q^2-\omega ^2/c_t^2\right) }}
%EndExpansion
\right) 
%TCIMACRO{\dfrac{4S_1S_3}{S^2} }
%BeginExpansion
{\displaystyle {4S_1S_3 \over S^2}}
%EndExpansion
\\ 
+\left( 
%TCIMACRO{
%\dfrac{q_2^2-\omega ^2/c_t^2}{\left( q^2-\omega ^2/c_t^2\right) } }
%BeginExpansion
{\displaystyle {q_2^2-\omega ^2/c_t^2 \over \left( q^2-\omega ^2/c_t^2\right) }}
%EndExpansion
+ 
%TCIMACRO{
%\dfrac{4\left( 1-\gamma ^2\right) q_1^2q_3^2}{\left( q^2-\omega ^2/c_t^2\right) \left( q^2-\omega ^2/c_l^2\right) } }
%BeginExpansion
{\displaystyle {4\left( 1-\gamma ^2\right) q_1^2q_3^2 \over \left( q^2-\omega ^2/c_t^2\right) \left( q^2-\omega ^2/c_l^2\right) }}
%EndExpansion
\right) 
%TCIMACRO{\dfrac{S_1^2}{S^2} }
%BeginExpansion
{\displaystyle {S_1^2 \over S^2}}
%EndExpansion
+%
%TCIMACRO{\dfrac{[H_{\circ }]^2}{\mu S^2T_{\circ }C_v\chi } }
%BeginExpansion
{\displaystyle {[H_{\circ }]^2 \over \mu S^2T_{\circ }C_v\chi }}
%EndExpansion
%TCIMACRO{\dfrac{i\omega }{q^2-i\omega /\chi }}
%BeginExpansion
{\displaystyle {i\omega  \over q^2-i\omega /\chi }}
%EndExpansion
\end{array}
\right) 
%TCIMACRO{\dfrac{dq_3}{2\pi } }
%BeginExpansion
{\displaystyle {dq_3 \over 2\pi }}
%EndExpansion
.  \label{11}
\end{equation}

Integral from cross component $S_1S_3$ is equal to zero and we receive

\begin{eqnarray}
\left( \mu S^2g\right) ^{-1} &=&%
%TCIMACRO{\dfrac{S_1^2}{S^2} }
%BeginExpansion
{\displaystyle {S_1^2 \over S^2}}
%EndExpansion
\left( 
%TCIMACRO{
%\dfrac{q_2^2-\omega ^2/c_t^2}{\sqrt{q_{\Vert }^2-\omega ^2/c_t^2}} }
%BeginExpansion
{\displaystyle {q_2^2-\omega ^2/c_t^2 \over \sqrt{q_{\Vert }^2-\omega ^2/c_t^2}}}
%EndExpansion
+%
%TCIMACRO{\dfrac{4q_1^2}{\omega ^2/c_t^2} }
%BeginExpansion
{\displaystyle {4q_1^2 \over \omega ^2/c_t^2}}
%EndExpansion
\left( \sqrt{q_{\Vert }^2-\omega ^2/c_l^2}-\sqrt{q_{\Vert }^2-\omega ^2/c_t^2%
}\right) \right)  \label{12} \\
&&+4%
%TCIMACRO{\dfrac{S_3^2}{S^2} }
%BeginExpansion
{\displaystyle {S_3^2 \over S^2}}
%EndExpansion
\left( 
%TCIMACRO{
%\dfrac{q_{\Vert }^2\left( q_{\Vert }^2-\omega ^2/c_t^2\right) }{\omega ^2/c_t^2} }
%BeginExpansion
{\displaystyle {q_{\Vert }^2\left( q_{\Vert }^2-\omega ^2/c_t^2\right)  \over \omega ^2/c_t^2}}
%EndExpansion
\left( 
%TCIMACRO{\dfrac 1{\sqrt{q_{\parallel }^2-\omega ^2/c_t^2}} }
%BeginExpansion
{\displaystyle {1 \over \sqrt{q_{\parallel }^2-\omega ^2/c_t^2}}}
%EndExpansion
-%
%TCIMACRO{\dfrac 1{\sqrt{q_{\Vert }^2-\omega ^2/c_l^2}} }
%BeginExpansion
{\displaystyle {1 \over \sqrt{q_{\Vert }^2-\omega ^2/c_l^2}}}
%EndExpansion
\right) -%
%TCIMACRO{\dfrac{\omega ^2/4c_t^2}{\sqrt{q_{\Vert }^2-\omega ^2/c_l^2}} }
%BeginExpansion
{\displaystyle {\omega ^2/4c_t^2 \over \sqrt{q_{\Vert }^2-\omega ^2/c_l^2}}}
%EndExpansion
\right)  \nonumber \\
&&+ 
%TCIMACRO{
%\dfrac{i\omega \delta /\chi }{\sqrt{q_{\Vert }^4+\omega ^2/\chi ^2}} }
%BeginExpansion
{\displaystyle {i\omega \delta /\chi  \over \sqrt{q_{\Vert }^4+\omega ^2/\chi ^2}}}
%EndExpansion
\left( \sqrt{%
%TCIMACRO{
%\dfrac{\sqrt{q_{\parallel }^4+\omega ^2/\chi ^2}+q_{\parallel }^2}2 }
%BeginExpansion
{\displaystyle {\sqrt{q_{\parallel }^4+\omega ^2/\chi ^2}+q_{\parallel }^2 \over 2}}
%EndExpansion
}+i\sqrt{%
%TCIMACRO{\dfrac{\sqrt{q_{\Vert }^4+\omega ^2/\chi ^2}-q_{\Vert }^2}2 }
%BeginExpansion
{\displaystyle {\sqrt{q_{\Vert }^4+\omega ^2/\chi ^2}-q_{\Vert }^2 \over 2}}
%EndExpansion
}\right) ,  \nonumber
\end{eqnarray}

where $\delta =%
%TCIMACRO{\dfrac{[H_{\circ }]^2}{\mu S^2T_{\circ }C_v}}
%BeginExpansion
{\displaystyle {[H_{\circ }]^2 \over \mu S^2T_{\circ }C_v}}
%EndExpansion
$ factor describing specific losses of energy of unit of the boundary area
on formation of a new phase at its movement.

If $\omega =\xi c_tq_{\Vert }$ then we obtain the dispersion law of coherent
interphase boundary:

\begin{eqnarray}
&&\ \cos ^2\theta \left( 
%TCIMACRO{\dfrac{\sin ^2\varphi -\xi ^2}{\sqrt{1-\xi ^2}} }
%BeginExpansion
{\displaystyle {\sin ^2\varphi -\xi ^2 \over \sqrt{1-\xi ^2}}}
%EndExpansion
+%
%TCIMACRO{\dfrac{4\cos ^2\varphi }{\xi ^2} }
%BeginExpansion
{\displaystyle {4\cos ^2\varphi  \over \xi ^2}}
%EndExpansion
\left( \sqrt{1-\gamma ^2\xi ^2}-\sqrt{1-\xi ^2}\right) \right)  \nonumber
\label{13} \\
&&\ +\sin ^2\theta \sqrt{%
%TCIMACRO{\dfrac{1-\xi ^2}{1-\gamma ^2\xi ^2} }
%BeginExpansion
{\displaystyle {1-\xi ^2 \over 1-\gamma ^2\xi ^2}}
%EndExpansion
}\left( 
%TCIMACRO{\dfrac{-\xi ^2}{\sqrt{1-\xi ^2}} }
%BeginExpansion
{\displaystyle {-\xi ^2 \over \sqrt{1-\xi ^2}}}
%EndExpansion
+%
%TCIMACRO{\dfrac 4{\xi ^2} }
%BeginExpansion
{\displaystyle {4 \over \xi ^2}}
%EndExpansion
\left( \sqrt{1-\gamma ^2\xi ^2}-\sqrt{1-\xi ^2}\right) \right)  \nonumber \\
\ &=& 
%TCIMACRO{
%\dfrac{\delta \xi c_t/\chi q_{\Vert }}{\sqrt{1+\xi ^2\left( c_t/\chi q_{\Vert }\right) ^2}} }
%BeginExpansion
{\displaystyle {\delta \xi c_t/\chi q_{\Vert } \over \sqrt{1+\xi ^2\left( c_t/\chi q_{\Vert }\right) ^2}}}
%EndExpansion
\left( \sqrt{%
%TCIMACRO{
%\dfrac{\sqrt{1+\xi ^2\left( c_t/\chi q_{\parallel }\right) ^2}-1}2 }
%BeginExpansion
{\displaystyle {\sqrt{1+\xi ^2\left( c_t/\chi q_{\parallel }\right) ^2}-1 \over 2}}
%EndExpansion
}+i\sqrt{%
%TCIMACRO{
%\dfrac{\sqrt{1+\xi ^2\left( c_t/\chi q_{\parallel }\right) ^2}+1}2 }
%BeginExpansion
{\displaystyle {\sqrt{1+\xi ^2\left( c_t/\chi q_{\parallel }\right) ^2}+1 \over 2}}
%EndExpansion
}\right) ,
\end{eqnarray}

where $\theta =\arctan 
%TCIMACRO{\dfrac{S_y}{S_x}}
%BeginExpansion
{\displaystyle {S_y \over S_x}}
%EndExpansion
$ is the dilatation angle, $\varphi =\arctan 
%TCIMACRO{\dfrac{q_y}{q_x}}
%BeginExpansion
{\displaystyle {q_y \over q_x}}
%EndExpansion
$ direction of wave distribution located near to a boundary.

Basic feature of moving coherent interfaces is the obligatory radiation or
absorption by it of the latent heat of phase transformation $[H_{\circ }]$.
In this connection the most essential channel of energy losses of border
oscillation is becomes dissipations caused by absorption of latent heat.
That allows to neglect attenuation, caused by viscosity, heat conductivity
of a crystal etc., considered in all earlier works on this theme.

Let's consider dislocations, moving in a crystal with constant speed $V$.
Then elastic stress created by it is possible to present as

\[
\sigma _{ik}^{ext}({\bf r},t)=\int \frac{d{\bf q}}{(2\pi )^3}\sigma
_{ik}^{ext}e^{i{\bf qr}-i\Omega t}, 
\]

where $\Omega ={\bf Vq}$. In view of this expression, for dissipations of
energy (1) we shall receive

\begin{equation}
\overline{D}=\int \frac{d{\bf q}_{\Vert }}{\left( 2\pi \right) ^2}\Omega
\left| f^{ext}({\bf q}_{\Vert })\right| ^2\text{Im }g({\bf q}_{\Vert }{\bf ,}%
\omega ).  \label{14}
\end{equation}

Taking into account, that on structure of created sheared stress of
considered defects their interaction is possible only by means of transverse
elastic fields, that for the given orientation of coherent interface from
every possible orientations of a dislocations line for us essential appear:

a) for screw dislocations

\begin{equation}
\sigma _{xz}=\frac{\mu b}{2\pi }\frac{y-Vt}{(y-Vt)^2+z_{\circ }^2}
\label{15}
\end{equation}

if ${\bf b}\Vert Ox$, ${\bf \tau }\Vert Ox$, where ${\bf \tau }$ is unit
vector tangent to dislocation line,

${\bf b}$ the Burgers vector of the dislocation,

$z_{\circ }$ distance up to boundary.

b) for edge dislocations

\begin{equation}
\sigma _{xz}=\frac{\mu b}{2\pi (1-\nu )}\frac{\left( x-V_xt\right) \left(
(x-V_xt)^2-z_{\circ }^2\right) }{\left( (x-V_xt)^2+z_{\circ }^2\right) ^2}
\label{16}
\end{equation}

if ${\bf b}\Vert Ox$, ${\bf \tau }\Vert Oy.$

Let's consider these cases separately.

\subsection{Screw dislocations}

Substituting a Fourier-image (15):

\[
\sigma _{xz}({\bf q}_{\Vert },\omega )=i(2\pi )^2\frac{\mu b}2\delta
(q_x)\delta (\omega -q_yV)e^{-\left| q_y\right| z_{\circ }} 
\]

into expression for dissipation of energy (14), with the account (2) we
shall receive

\begin{eqnarray}
D &=&(2\pi )^2\left( \frac{\mu b}2\right) ^2S_1^2\int dq_yq_yV\text{ Im }%
g(0,q_y,Vq_y)e^{-2\left| q_y\right| z_{\circ }}  \label{17} \\
\ &=&\left( \pi bV\cos \theta \right) ^2\frac{\delta \mu }\chi \int \frac{%
%TCIMACRO{\dfrac{dq_y}{q_y} }
%BeginExpansion
{\displaystyle {dq_y \over q_y}}
%EndExpansion
\sqrt{%
%TCIMACRO{
%\dfrac{\sqrt{1+V^2/q_y^2\chi ^2}+1}{2\left( 1+V^2/q_y^2\chi ^2\right) } }
%BeginExpansion
{\displaystyle {\sqrt{1+V^2/q_y^2\chi ^2}+1 \over 2\left( 1+V^2/q_y^2\chi ^2\right) }}
%EndExpansion
}e^{-2\left| q_y\right| z_{\circ }}}{\left( A\cos ^2\theta +4B\sin ^2\theta -%
%TCIMACRO{\dfrac{V\delta }{\chi q_y} }
%BeginExpansion
{\displaystyle {V\delta  \over \chi q_y}}
%EndExpansion
\sqrt{%
%TCIMACRO{
%\dfrac{\sqrt{1+V^2/q_y^2\chi ^2}-1}{2\left( 1+V^2/q_y^2\chi ^2\right) } }
%BeginExpansion
{\displaystyle {\sqrt{1+V^2/q_y^2\chi ^2}-1 \over 2\left( 1+V^2/q_y^2\chi ^2\right) }}
%EndExpansion
}\right) ^2+\left( 
%TCIMACRO{\dfrac{V\delta }{\chi q_y} }
%BeginExpansion
{\displaystyle {V\delta  \over \chi q_y}}
%EndExpansion
\right) ^2 
%TCIMACRO{
%\dfrac{\sqrt{1+V^2/q_y^2\chi ^2}+1}{2\left( 1+V^2/q_y^2\chi ^2\right) } }
%BeginExpansion
{\displaystyle {\sqrt{1+V^2/q_y^2\chi ^2}+1 \over 2\left( 1+V^2/q_y^2\chi ^2\right) }}
%EndExpansion
}  \nonumber
\end{eqnarray}

where 
\[
A=\sqrt{1-V^2/c_t^2}, 
\]

\[
B=%
%TCIMACRO{\dfrac{\left( 1-V^2/c_t^2\right) }{V^2/c_t^2}}
%BeginExpansion
{\displaystyle {\left( 1-V^2/c_t^2\right)  \over V^2/c_t^2}}
%EndExpansion
\left( 
%TCIMACRO{\dfrac 1{\sqrt{1-V^2/c_t^2}}}
%BeginExpansion
{\displaystyle {1 \over \sqrt{1-V^2/c_t^2}}}
%EndExpansion
-%
%TCIMACRO{\dfrac 1{\sqrt{1-V^2/c_l^2}}}
%BeginExpansion
{\displaystyle {1 \over \sqrt{1-V^2/c_l^2}}}
%EndExpansion
\right) -%
%TCIMACRO{\dfrac{V^2/4c_t^2}{\sqrt{1-V^2/c_l^2}}}
%BeginExpansion
{\displaystyle {V^2/4c_t^2 \over \sqrt{1-V^2/c_l^2}}}
%EndExpansion
. 
\]

Results of the numerical analysis of integral in (17) for concrete meanings $%
\gamma ^2=0.3$ and $\theta =\pi /4$ show on figure. For screw dislocation,
driven perpendicularly to vector of shift $S$ the maximum is necessary on $%
V^2/c_t^2\simeq 0.914$, i.e. near to a sound meaning of speed of movement
dislocation.

\subsection{Edge dislocations}

Driven along an axis $Ox$ edge dislocations creates around of itself
chopping off stress (16)

\[
\sigma _{xz}({\bf q}_{\Vert },\omega )=i(2\pi )^2\frac{\mu b}{2(1-\nu )}%
\delta (q_y)\delta (\omega -q_xV)e^{-\left| q_y\right| z_{\circ
}}(1-q_xz_{\circ }), 
\]

by means of which there is its interaction with coherent boundary and
outflow of energy on its excitation. In this case for dissipation of energy
D of unit of length of dislocations shall receive:

\begin{equation}
\overline{D}=\left( \frac{\pi bV\cos \theta }{(1-\nu )}\right) ^2%
%TCIMACRO{\dfrac{\delta \mu }\chi }
%BeginExpansion
{\displaystyle {\delta \mu  \over \chi}}
%EndExpansion
%TCIMACRO{\dint }
%BeginExpansion
\displaystyle \int 
%EndExpansion
\frac{%
%TCIMACRO{\dfrac{dq_x}{q_x} }
%BeginExpansion
{\displaystyle {dq_x \over q_x}}
%EndExpansion
\sqrt{%
%TCIMACRO{
%\dfrac{\sqrt{1+V^2/q_x^2\chi ^2}+1}{2\left( 1+V^2/q_x^2\chi ^2\right) } }
%BeginExpansion
{\displaystyle {\sqrt{1+V^2/q_x^2\chi ^2}+1 \over 2\left( 1+V^2/q_x^2\chi ^2\right) }}
%EndExpansion
}e^{-2\left| q_x\right| z_{\circ }}(1-q_xz_{\circ })^2}{\left( A\cos
^2\theta +4B\sin ^2\theta -%
%TCIMACRO{\dfrac{V\delta }{\chi q_x} }
%BeginExpansion
{\displaystyle {V\delta  \over \chi q_x}}
%EndExpansion
\sqrt{%
%TCIMACRO{
%\dfrac{\sqrt{1+V^2/q_x^2\chi ^2}-1}{2\left( 1+V^2/q_x^2\chi ^2\right) } }
%BeginExpansion
{\displaystyle {\sqrt{1+V^2/q_x^2\chi ^2}-1 \over 2\left( 1+V^2/q_x^2\chi ^2\right) }}
%EndExpansion
}\right) ^2+\left( 
%TCIMACRO{\dfrac{V\delta }{\chi q_x} }
%BeginExpansion
{\displaystyle {V\delta  \over \chi q_x}}
%EndExpansion
\right) ^2 
%TCIMACRO{
%\dfrac{\sqrt{1+V^2/q_x^2\chi ^2}+1}{2\left( 1+V^2/q_x^2\chi ^2\right) } }
%BeginExpansion
{\displaystyle {\sqrt{1+V^2/q_x^2\chi ^2}+1 \over 2\left( 1+V^2/q_x^2\chi ^2\right) }}
%EndExpansion
},  \label{18}
\end{equation}

where

\[
A=%
%TCIMACRO{\dfrac 4{V^2/c_t^2}}
%BeginExpansion
{\displaystyle {4 \over V^2/c_t^2}}
%EndExpansion
\left( \sqrt{1-V^2/c_l^2}-\sqrt{1-V^2/c_t^2}\right) -%
%TCIMACRO{\dfrac{V^2/c_t^2}{\sqrt{1-V^2/c_t^2}}}
%BeginExpansion
{\displaystyle {V^2/c_t^2 \over \sqrt{1-V^2/c_t^2}}}
%EndExpansion
, 
\]

and B coincides with (17). The numerical analysis (18) qualitatively does
not differ from results received for screw dislocations (17). However, here
it is necessary to note displacement of a maximum of dynamic braking in the
party of lower speeds, though and all the same, close to sound $%
V^2/c_t^2\simeq 0.985$.

The transition from (7) - (8) to usually measure experimentally size -
factor of dynamic dislocation braking - is made by a standard equation:

\[
B=D/V^2. 
\]

In summary we shall note, considered above mechanisms dissipation of energy,
it is necessary to take into account obviously, and in crystals with more
complex structure: ferroelectrics, ferroelectrics-ferroelastics etc.

\newpage

\begin{center}
Figure.
\end{center}

Dependence of dissipation of energy $D$ rectilinear screw dislocation from
speed $V$ and factor $\alpha $, which describes losses of boundaries energy
on formation of a new phase at her movement.

\end{document}